\documentclass[preprint, superscriptaddress, preprintnumbers, amsmath, amssymb]{revtex4}

\allowdisplaybreaks
\allowdisplaybreaks[4]
\usepackage{CJK}
\usepackage{graphicx}
\usepackage{dcolumn}
\usepackage{bm}
\usepackage[dvipdfm,
            pdfstartview=FitH,
            CJKbookmarks=true,
            bookmarksnumbered=true,
            bookmarksopen=true,
            colorlinks=true,
            pdfborder=001,
            linkcolor=blue,
            anchorcolor=blue,
            citecolor=blue
            ]{hyperref}

\begin{document}

\begin{CJK}{GBK}{song}

\title{Behavior of the collective rotor in wobbling motion}

\author{E. Streck}
\affiliation{Physik-Department, Technische Universit\"{a}t
M\"{u}nchen, D-85747 Garching, Germany}

\author{Q. B. Chen}\email{qbchen@pku.edu.cn}
\affiliation{Physik-Department, Technische Universit\"{a}t
M\"{u}nchen, D-85747 Garching, Germany}

\author{N. Kaiser}\email{nkaiser@ph.tum.de}
\affiliation{Physik-Department, Technische Universit\"{a}t
M\"{u}nchen, D-85747 Garching, Germany}

\author{Ulf-G. Mei{\ss}ner}\email{meissner@hiskp.uni-bonn.de}
\affiliation{Helmholtz-Institut f\"{u}r Strahlen- und Kernphysik and
Bethe Center for Theoretical Physics, Universit\"{a}t Bonn, D-53115
Bonn, Germany}

\affiliation{Institute for Advanced Simulation, Institut f\"{u}r
Kernphysik, J\"{u}lich Center for Hadron Physics and JARA-HPC,
Forschungszentrum J\"{u}lich, D-52425 J\"{u}lich, Germany}

\date{\today}

\begin{abstract}
The behavior of the collective rotor in  wobbling motion is investigated
within the particle-rotor model for the nucleus $^{135}$Pr by transforming
the wave functions from the $K$-representation to the $R$-representation.
After reproducing the experimental energy spectra and wobbling
frequencies, the evolution of the wobbling mode in $^{135}$Pr, from transverse
at low spins to longitudinal at high spins, is illustrated by
the distributions of the total angular momentum in the intrinsic reference
frame (azimuthal plot). Finally, the coupling schemes of the angular
momenta of the rotor and the high-$j$ particle for transverse and
longitudinal wobbling are obtained from the analysis of the probability
distributions of the rotor angular momentum ($R$-plots) and their projections
onto the three principal axes ($K_R$-plots).

\end{abstract}

\maketitle


\section{Introduction}

As a quantum-mechanical complex many-body system, an atomic nucleus can
possess a wide variety of shapes in its ground and excited states.
The shapes may range from spherical to deformed (quadrupole, octupole, etc.)
and even more exotic shapes, such as superdeformed and
tetrahedral are possible. At the same time, the atomic nucleus can exhibit various modes of
collective excitations. Obviously, the modes of collective motion are strongly
correlated with the nuclear shapes. For example, only a nucleus with triaxial
deformation can possibly have chiral rotation~\cite{Frauendorf1997NPA} or
wobbling motion~\cite{Bohr1975}.

The wobbling motion was first proposed by Bohr and Mottelson in the
1970s~\cite{Bohr1975}. It occurs in the case when the rotation of a
triaxial nucleus about the principal axis with the largest moment of inertia
(MoI) is quantum mechanically disturbed by rotations about the other two
principal axes, and hence it precesses and wobbles around the axis with the largest MoI.
The energy spectra related to the wobbling motion are called wobbling bands,
and these consist of sequences of $\Delta I = 2$ rotational bands built on different
wobbling-phonon excitations~\cite{Bohr1975}.

The excitation spectrum of the wobbling motion is characterized by the
wobbling frequency. For the originally predicted wobbler (a triaxial
rotor built up by an even-even nucleus)~\cite{Bohr1975}, the wobbling frequency
increases with spin. For an odd-mass nucleus, the triaxial rotor is
coupled with a high-$j$ quasiparticle, and in this case two different wobbling
modes were proposed by Frauendorf and D\"{o}nau~\cite{Frauendorf2014PRC}. One of
them is called \emph{longitudinal} wobbling, in which the quasiparticle
angular momentum is parallel to the principal axis with the largest
MoI. The other one is named \emph{transverse} wobbling, since the quasiparticle
angular momentum is perpendicular to the principal axis with the
largest MoI. According to Ref.~\cite{Frauendorf2014PRC}, the wobbling frequency
of a longitudinal wobbler increases, while that of a transverse wobbler
decreases with increasing spin.

Wobbling bands have been reported in the mass region $A\approx 160$
for the isotopes $^{161}$Lu~\cite{Bringel2005EPJA},
$^{163}$Lu~\cite{Odegaard2001PRL, Jensen2002PRL},
$^{165}$Lu~\cite{Schonwasser2003PLB}, $^{167}$Lu~\cite{Amro2003PLB},
and $^{167}$Ta~\cite{Hartley2009PRC}, in the mass region $A\approx
110$ for $^{112}$Ru~\cite{S.J.Zhu2009IJMPE} and
$^{114}$Pd~\cite{Y.X.Luo2013proceeding}, and recently in the mass
region $A\approx 130$ for $^{135}$Pr~\cite{Matta2015PRL} and
$^{133}$La~\cite{Biswas2017arXiv}.

Interestingly, the isotope $^{135}$Pr does not only possess the transverse wobbling mode,
but also exhibits a transition from transverse to longitudinal wobbling~\cite{Matta2015PRL}.
Hence, $^{135}$Pr is an excellent candidate for understanding the wobbling motion and
has attracted a lot of theoretical attention. These studies can be briefly
summarized as follows.
\begin{itemize}
  \item [(i)] In Refs.~\cite{Frauendorf2014PRC, Matta2015PRL}, the tilted axis
  cranking (TAC) model with the Strutinsky shell corrections and the particle-rotor model
  (PRM) were employed to confirm the wobbling nature of the experimental
  energy spectra and the electromagnetic transition probabilities.
  \item [(ii)] In Ref.~\cite{Sheikh2016PS}, the multi-quasiparticle triaxial projected
  shell model (TPSM) approach was used to extract the probabilities of various projected
  configurations in the wave functions of the yrast and the wobbling bands.
  \item [(iii)] In Ref.~\cite{Q.B.Chen2016PRC_v1}, a collective Hamiltonian
  method based on the TAC approach was applied to reveal the microscopic
  mechanisms underlying the variation of the wobbling frequency with spin and
  the transition from transverse to longitudinal wobbling.
  \item [(iv)] In Ref.~\cite{Tanabe2017PRC}, the Holstein-Primakoff boson expansion
  was applied to the PRM to examine the stability of the wobbling motion.
  \item [(v)] In Ref.~\cite{Budaca2018PRC}, a time-dependent variational
  method, with coherent angular momentum states as variational states,
  was adopted to treat the PRM (specialized to a high-$j$ quasiparticle aligned
  rigidly with one principal axis) and to obtain analytical solutions for the energy
  spectra and electromagnetic transition probabilities.
\end{itemize}

However, still no attempt has been made to investigate the detailed structure
of wave functions of the collective rotor in wobbling bands. Taking $^{135}$Pr
as an example, we investigate in this paper the behavior of the collective rotor
angular momentum in wobbling motion using the PRM.

For this purpose, one has to express the PRM wave function in terms of
the weak-coupling basis~\cite{Bohr1975, Ring1980book}, in which both $R$ (rotor
angular momentum quantum number) and $K_R$ (projection on a principal axis) are
good quantum numbers. This transformation gives the $R$-representation.
From the corresponding probability distributions one can derive (by summation)
the {\it $R$-plot} and the {\it $K_R$-plots}.

Usually, the PRM wave functions are formulated in terms of the strong
coupling basis~\cite{Bohr1975, Ring1980book}, where the projection
of the total spin onto the 3-axis of the intrinsic frame is a
good quantum number, denoted by $K$. In this $K$-representation,
$R$ and $K_R$ do not appear explicitly. Therefore, in order to obtain the
$R$-plot and the $K_R$-plots, one has to transform the PRM wave function from the
$K$-representation to the $R$-representation. This technique has been applied
for a long time to take into account $R$-dependent MoIs~\cite{Bohr1975,
Faessler1975PLB, Toki1976PLB, Smith1976PRC, Ragnarsson1988HI,
Mukherjee1994PRC, Modi2017PRC} or shape fluctuations
of the rotor~\cite{Donau1977PLB, Q.Suan2017PRC}
in the description of rotational spectrum, or to calculate decay widths
of proton emitters~\cite{Esbensen2000PRC, Davids2004PRC,
Modi2017PRC_v1, Modi2017PRC_v2}. The probability distributions of the
rotor angular momentum were also obtained before in an
analysis of rotational spectra of axially symmetric nuclei~\cite{Smith1976PRC}.
Here, it is employed for the first time to investigate the detailed wave
function structure of the collective rotor in the wobbling motion
of a triaxial nucleus.


\section{Theoretical framework}\label{sec1}

\subsection{Particle-rotor Hamiltonian}

The total Hamiltonian of the PRM takes the form~\cite{Bohr1975, Ring1980book}
\begin{align}\label{eq1}
\hat{H}_\textrm{PRM}=\hat{H}_{\rm coll}+\hat{H}_{\rm intr},
\end{align}
with $\hat{H}_{\textrm{coll}}$ the collective rotor Hamiltonian
\begin{align}\label{eq6}
\hat{H}_{\rm coll}
 &=\sum_{k=1}^3 \frac{\hat{R}_k^2}{2\mathcal{J}_k}\\
\label{eq11}
 &=\sum_{k=1}^3 \frac{(\hat{I}_k-\hat{j}_k)^2}{2\mathcal{J}_k},
\end{align}
where the index $k=1$, 2, 3 denotes the three principal axes of the body-fixed frame.
Here, $\hat{R}_k$ and $\hat{I}_k$ are the angular momentum operators
of the collective rotor and the total nucleus, and
$\hat{j}_k$ is the angular momentum operator of a valence nucleon.
Moreover, the parameters $\mathcal{J}_k$ are the three principal
MoIs. When calculating  matrix elements
of $\hat{H}_{\rm coll}$, the $R$-representation is most conveniently
used for its form in Eq.~(\ref{eq6}), while Eq.~(\ref{eq11}) is preferable
in the $K$-representation.

The intrinsic Hamiltonian $\hat{H}_{\rm intr}$ describes a single
valence nucleon in a high-$j$ shell
\begin{align}\label{eq2}
 \hat{H}_{\rm intr}=\pm \frac{1}{2}C\bigg\{\cos \gamma\bigg(\hat{j}_3^2-\frac{j(j+1)}{3}\bigg)
                +\frac{\sin \gamma}{2\sqrt{3}}\big(\hat{j}_+^2+\hat{j}_-^2\big)\bigg\},
\end{align}
where $\pm$ refers to a particle or a hole state. The angle $\gamma$
serves as the triaxial deformation parameter and the coefficient $C$
is proportional to the quadrupole deformation parameter $\beta$. We
take in the present work the same form of $C$ as in Ref.~\cite{S.Y.Wang2009CPL}.

\subsection{Basis transformation from $K$-representation to $R$-representation}

As mentioned in the Introduction, the PRM Hamiltonian (\ref{eq1}) is usually
solved by diagonalization in the strong-coupling basis
($K$-representation)~\cite{Bohr1975, Ring1980book}
\begin{align}\label{eq12}
 |IMKj\Omega\rangle
  =\sqrt{\frac{2I+1}{16\pi^2}}\Big[D_{MK}^I(\bm{\omega})|j\Omega\rangle
 +(-1)^{I-j}D_{M-K}(\bm{\omega})|j-\Omega\rangle \Big],
\end{align}
where $I$ denotes the total angular momentum quantum number of
the odd-mass nuclear system (rotor plus particle) and $M$ is the
projection onto the 3-axis of the laboratory frame. Furthermore,
$\Omega$ is the 3-axis component of the particle angular momentum $j$
in the intrinsic frame, and $D_{MK}^I(\bm{\omega})$ are the usual
Wigner-functions, depending on the three Euler angles $\bm{\omega}=(\psi^\prime,
\theta^\prime,\phi^\prime)$. Under the requirement of the $\textrm{D}_2$ symmetry of a triaxial
nucleus~\cite{Bohr1975}, $K$ and $\Omega$ take the values:
$K=-I,\dots,I$, $\Omega=-j, \dots,j$, $K-\Omega \geq 0$ and even; and if $K-\Omega=0$,
$K=\Omega>0$.

As seen in the $K$-representation (\ref{eq12}), the rotor angular momentum $R$ does
not appear explicitly. In order to obtain the wave function of the rotor
in the $R$-representation, one has to transform the basis. The details of
this transformation can be found in Refs.~\cite{Davids2004PRC, Modi2017PRC}. Here,
we outline the main ingredients.

The wave function of the total nuclear system in the laboratory frame
can be expressed in the $R$-representation as
\begin{align}\label{eq5}
 |IMjR\tau\rangle=\sum_{m,M_R}\langle j m R M_R|IM\rangle~|jm\rangle \otimes |RM_R \tau \rangle,
\end{align}
where $m$ and $M_R$ are the projections of $j$ and $R$ on the 3-axis of the laboratory
frame. Obviously, the appearance of Clebsch-Gordan coefficients requires
$M=m+M_R$, and the values of $R$ must satisfy the triangular condition
$|I-j| \leq R \leq I+j$ of angular momentum coupling. At the moment,
the additional quantum number $\tau$, related to the projection
of $R$ on a body-fixed axis, is not yet specified. Now we perform the
transformation from the $R$-representation to the $K$-representation.

In the $K$-representation, the quantum number $\tau$ is
identified with the projection $K_R$ of $R$ on a principal axis. Making use of
Wigner-functions, the wave functions of the particle and the rotor in Eq.~(\ref{eq5})
can be written as
\begin{align}
\label{eq3}
 |jm\rangle &=\sum_{\Omega=-j}^j D_{m\Omega}^j(\bm{\omega})|j\Omega\rangle,\\
\label{eq4}
 |RM_R K_R\rangle &=\sqrt{\frac{2R+1}{16\pi^2(1+\delta_{K_R0})}}
  \Big[D_{M_RK_R}^R(\bm{\omega})+(-1)^R D_{M_R-K_R}^R(\bm{\omega})\Big],
\end{align}
where $K_R$ is an even integer ranging from $0$ to $R$, with $K_R=0$
is excluded for odd $R$. Both restrictions come from the $\textrm{D}_2$
symmetry of a triaxial nucleus~\cite{Bohr1975}. Note that
for an axially symmetric nucleus, $R$ can only take even integer values
since $K_R$ must be zero.

Substituting Eqs.~(\ref{eq3}) and (\ref{eq4}) into Eq.~(\ref{eq5}), one obtains
\begin{align}\label{eq13}
 |IMjRK_R\rangle =\sum_{K,\Omega}A_{j\Omega,RK_R}^{IK}|IMKj\Omega\rangle,
\end{align}
with the expansion coefficients
\begin{align}
 A_{j\Omega,RK_R}^{IK}=\sqrt{\frac{2R+1}{2I+1}}
  \langle j\Omega RK_R|IK\rangle \sqrt{1+\delta_{K_R0}},
\end{align}
determined by Clebsch-Gordan coefficients (hence $K=K_R+\Omega$).

Obviously, the transformation between the $K$-representation and
the $R$-representation is an orthogonal transformation, and therefore the
expansion coefficients satisfy
\begin{align}
 \sum_{K,\Omega} A_{j\Omega,RK_R}^{IK} A_{j\Omega,R^\prime K_{R}^\prime}^{IK}
  &=\delta_{RR^\prime}\delta_{K_R K_{R}^\prime},\\
 \sum_{R,K_R} A_{j\Omega,RK_R}^{IK} A_{j\Omega^\prime,RK_R}^{IK^\prime}
  &=\delta_{\Omega\Omega^\prime}\delta_{KK^\prime}.
\end{align}
Due to the orthogonality property, the inverse transformation
follows immediately as
\begin{align}\label{eq14}
 |IMKj\Omega\rangle=\sum_{R,K_R} A_{j\Omega,RK_R}^{IK} |IMjRK_R\rangle.
\end{align}
To this end, we have successfully transformed the PRM basis functions
from the $K$-representation to $R$-representation.

Eq. (\ref{eq14}) allows us also to calculate the matrix elements
of the collective rotor Hamiltonian in the $K$-representation as
\begin{align}
 &\quad \langle IMK^\prime j\Omega^\prime |\hat{H}_{\textrm{coll}}|IMKj\Omega\rangle\notag\\
 &=\sum_{R,K_R,K_R^\prime} A_{j\Omega^\prime,RK_R^\prime}^{IK^\prime}
 \langle IMjRK_R^\prime|\hat{H}_{\textrm{coll}}|IMjRK_R \rangle
 A_{j\Omega,RK_R}^{IK}\notag\\
 &=\sum_{R,K_R,K_R^\prime} A_{j\Omega^\prime,RK_R^\prime}^{IK^\prime}
 \Big(\sum_i c_{K_R^\prime}^{Ri} E_{Ri} c_{K_R}^{Ri}\Big)
 A_{j\Omega,RK_R}^{IK},
\end{align}
where the energies $E_{Ri}$ and corresponding expansion coefficients
$c_{K_R}^{Ri}$ ($i$ labels the different eigenstates) are obtained by
diagonalizing the collective rotor Hamiltonian $\hat{H}_{\textrm{coll}}$
in the basis $|RM_RK_R\rangle$ introduced in Eq.~(\ref{eq4})
\begin{align}
 \hat{H}_{\textrm{coll}}|RM_Ri\rangle &= E_{Ri}|RM_Ri\rangle,\\
 |RM_Ri\rangle &=\sum_{K_R} c_{K_R}^{Ri} |RM_RK_R\rangle.
\end{align}
In such a calculation, $R$-dependent MoIs can be easily
implemented in the PRM to obtain a better description of high spin
states~\cite{Bohr1975, Faessler1975PLB, Toki1976PLB, Smith1976PRC,
Ragnarsson1988HI, Modi2017PRC}. The main focus of the present work is
on the probability distributions of the rotor angular momentum
derived from the transformation (\ref{eq14}).

\subsection{$R$-plot and $K_R$-plot}

With the above preparations, the PRM eigenfunctions can be expressed as
\begin{align}\label{eq7}
 |IM\rangle &=\sum_{K,\Omega} d_{K,\Omega} |IMKj\Omega\rangle\\
  &=\sum_{K,\Omega} d_{K,\Omega} \sum_{R,K_R} A_{j\Omega,RK_R}^{IK}
   \sum_{m,M_R}\langle j m R M_R|IM\rangle |RM_R K_R \rangle |jm\rangle,
\end{align}
where the (real) expansion coefficients $d_{K,\Omega}$ are obtained by solving the total
PRM Hamiltonian $\hat{H}_{\rm coll}+\hat{H}_{\rm intr}$ in Eq.~(\ref{eq1}).
Hence, the probabilities for given $R$ and $K_R$ are calculated as
\begin{align}
 P_{R,K_R}=\Big(\sum_{K,\Omega} d_{K,\Omega} A_{j\Omega,RK_R}^{IK}\Big)^2,
\end{align}
and they satisfy the normalization condition
\begin{align}
 \sum_{R,K_R}P_{R,K_R}=1.
\end{align}
The $R$-plot consists of the summed probabilities
\begin{align}\label{eq9}
 P_R=\sum_{K_R} P_{R,K_R},
\end{align}
whereas in the $K_R$-plot the probabilities are summed differently
\begin{align}\label{eq10}
 P_{K_R}=\sum_R P_{R,K_R}.
\end{align}
Moreover, the expectation value of the squared angular momentum operator
$\hat{R}_3^2$ follows as
\begin{align}
 \langle IM|\hat{R}_3^2|IM\rangle=\sum_{R,K_R} K_R^2 P_{R,K_R}.
\end{align}

\subsection{Azimuthal plot}

In this work, we want to illustrate the angular momentum geometry
of the wobbling motion by a profile on the $(\theta, \varphi)$
unit-sphere, called azimuthal plot~\cite{F.Q.Chen2017PRC, Q.B.Chen2018arXiv}.
Here, $(\theta,\varphi)$ are the orientation angles of the angular momentum
vector $\bm{I}$ (expectation value with $M=I$) with respect to the intrinsic frame.
The polar angle $\theta$ is the angle between $\bm{I}$
and the $3$-axis, whereas the azimuthal angle $\varphi$ is the
angle between the projection of $\bm{I}$ on the 12-plane and the 1-axis.
The profiles can be obtained by relating the orientation angles $(\theta,\varphi)$
to the Euler angles $\bm{\omega}=(\psi^\prime,\theta,
\pi-\varphi)$~\cite{Frauendorf1997NPA, F.Q.Chen2017PRC}, where
the $z$-axis in the laboratory frame is chosen along $\bm{I}$.
The profiles are calculated from the PRM eigenfunctions (\ref{eq7}) as
\begin{align}\label{eq15}
  \mathcal{P}(\theta,\varphi)
   &=2\pi \sum_{\Omega_p} \Big|\sum_{K,\Omega} d_{K,\Omega}
   \sqrt{\frac{2I+1}{16\pi^2}} \Big[D_{IK}^I(\psi^\prime,\theta,\pi-\varphi)
    \delta_{\Omega_p,\Omega}\notag\\
   &+(-1)^{I-j}D_{I-K}^I(\psi^\prime, \theta,\pi-\varphi)
  \delta_{\Omega_p,-\Omega}\Big]\Big|^2,
\end{align}
where the factor $2\pi$ comes from the integral over $\psi^\prime$.
Note that $D_{IK}^I(\psi^\prime,\theta^\prime,\varphi^\prime)\sim e^{-iI\psi^\prime}$,
and therefore the right side of Eq.~(\ref{eq15}) is $\psi^\prime$-independent.
The profiles $\mathcal{P}(\theta,\varphi)$ fulfil the normalization
condition
\begin{align}
 \int_{0}^\pi d\theta~\sin\theta \int_{-\pi}^{\pi} d\varphi~
 \mathcal{P}(\theta,\varphi)=1.
\end{align}
Due to the combination of Wigner functions required by the $\textrm{D}_2$ symmetry
in Eq.~(\ref{eq15}), $\mathcal{P}(\theta,\varphi)$ fulfils the following relations:
$\mathcal{P}(\theta,\varphi)=\mathcal{P}(\theta,-\varphi)=\mathcal{P}(\theta,\pi-\varphi)
=\mathcal{P}(\pi-\theta,\varphi)$. Therefore, the complete information is contained
in the angle ranges $0\leq \theta \leq  \pi/2$ and $0 \leq \varphi \leq \pi/2$.

\section{Numerical details}\label{sec2}

In our calculation of the wobbling bands in $^{135}$Pr, the configuration of
the proton is taken as $\pi(1h_{11/2})^1$. Following Refs.~\cite{Frauendorf2014PRC,
Matta2015PRL, Q.B.Chen2016PRC_v1}, the quadrupole deformation parameters of this
configuration have the values $\beta=0.17$ and $\gamma=-26.0^\circ$.
With this assignment of $\gamma$, the 1-, 2-, and 3-axes are the short ($s$-),
intermediate ($i$-), and long ($l$-) axes of the ellipsoid, respectively. The principal
MoIs are taken as $\mathcal{J}_1$, $\mathcal{J}_2$,
$\mathcal{J}_3$=13.0, 21.0, 4.0~$\hbar^2/\rm{MeV}$~\cite{Frauendorf2014PRC,
Q.B.Chen2016PRC_v1}. In this case, the $i$-axis is the axis with the
largest MoI.


\section{Results and discussion}\label{sec3}

\subsection{Energy spectra of wobbling bands}

\begin{figure}[!ht]
  \begin{center}
    \includegraphics[width=8 cm]{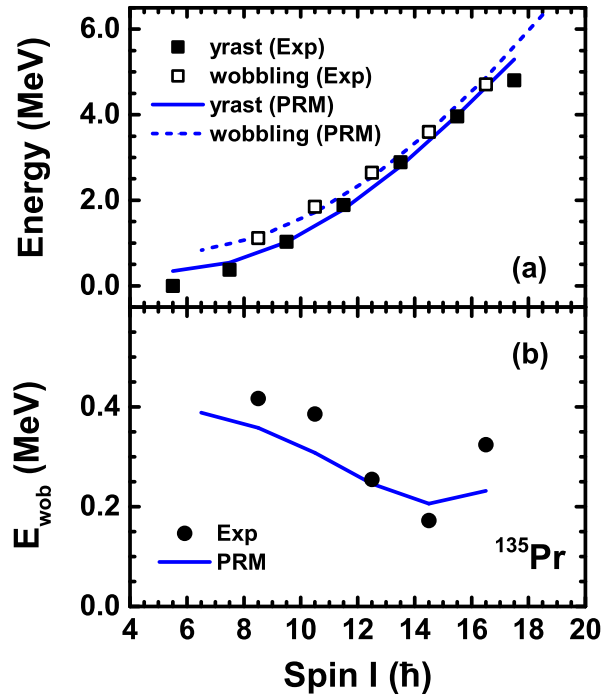}
    \caption{Energy spectra of the yrast (zero-phonon) and wobbling bands
    (one-phonon) (a) and the corresponding wobbling frequency (b) in $^{135}$Pr as
    functions of the total spin $I$ calculated in the PRM in comparison to the
    experimental data of Ref.~\cite{Matta2015PRL}.}\label{fig10}
  \end{center}
\end{figure}

In Fig.~\ref{fig10}(a), the energy spectra of the yrast and
wobbling bands calculated in the PRM are compared with the experimental
data~\cite{Matta2015PRL}. A similar figure has been given in
Ref.~\cite{Q.B.Chen2016PRC_v1}, where the collective Hamiltonian method
has been used. For both approaches, good agreement between the theoretical
calculations and the data can be obtained.

From the energy spectra, the wobbling frequencies $E_{\rm wob}(I)$ of the
theoretical calculation and the data are extracted (as differences) and shown in
Fig.~\ref{fig10}(b) as a function of spin $I$. In the region $I\leq 14.5\hbar$, both
the theoretical and experimental wobbling frequencies decrease with spin,
which provides evidence for transverse wobbling motion. At higher spin
($I\geq 14.5\hbar$), the experimental wobbling frequency shows
an increasing trend, which indicates that the wobbling mode changes from
transverse to longitudinal~\cite{Matta2015PRL}. The PRM
calculations can reproduce this transition well.

\begin{figure}[!ht]
  \begin{center}
    \includegraphics[width=8 cm]{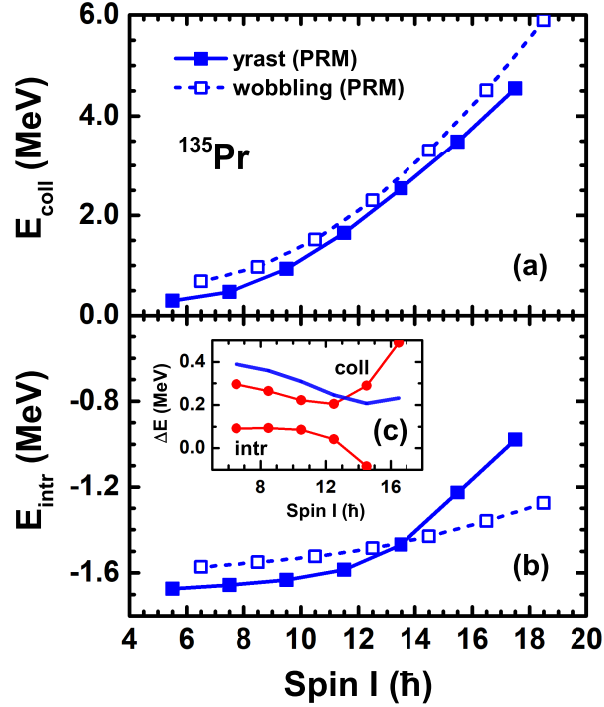}
    \caption{Energy expectation values of the collective rotor (a) and the valence proton (b)
    calculated in the PRM for the yrast and wobbling bands in $^{135}$Pr, together with
    their differences and the full wobbling frequency (c) as functions of the spin $I$.}\label{fig80}
  \end{center}
\end{figure}

In Fig.~\ref{fig80}, the separated energy expectation values $E_{\rm{coll}}$ and $E_{\rm{intr}}$
of the collective rotor Hamiltonian $\hat{H}_{\rm{coll}}$ and the intrinsic
single-proton Hamiltonian $\hat{H}_{\rm{intr}}$ as calculated in the PRM for the yrast
and wobbling bands in $^{135}$Pr are shown as functions of the spin $I$, together with
the differences $\Delta E$ in the two bands and the wobbling frequency.

It is seen that $E_{\rm{coll}}$ increases with the spin, and apparently, the
yrast band has lower $E_{\rm{coll}}$ than the wobbling band.
The difference of $E_{\rm{coll}}$ in the wobbling and yrast band
decreases up to $I=12.5\hbar$, and then it increases rapidly.

In the region $I\leq 11.5\hbar$, the values of $E_{\rm{intr}}$ in
the yrast and wobbling bands do not vary much, which implies that
the alignment of the proton particle along the $s$-axis remains
almost unchanged. This is a specific feature of the wobbling mode in
contrast to the cranking mode, where the alignment of the single
particle varies with the spin~\cite{Odegaard2001PRL,
Hamamoto2002PRC}. The values of $E_{\rm{intr}}$ in the yrast band
are a bit smaller than those in the wobbling band, but their
differences stay almost constant. As a consequence, the decrease of
the wobbling frequencies originates mainly from the decrease of the
$E_{\rm{coll}}$ differences.

However, from $I=13.5\hbar$ upward, $E_{\rm{intr}}$ of the yrast band
increases rapidly, which is caused by the change of alignment of the proton
particle from the $s$-axis towards the $i$-axis, driven by
the Coriolis interaction. As revealed by the azimuthal plots (discussed later),
this corresponds to a change of the rotational mode from along a principal axis
($s$-axis) to a planar rotation (with $\bm{I}$ lying in the $si$-plane).
This rearrangement leads to much larger values of $E_{\rm{intr}}$ in the yrast
band than in the wobbling band, and hence their difference
decreases to negative value for $I\geq 12.5\hbar$.

\subsection{Azimuthal plot}

The successful reproduction of the energy spectra in the yrast
and wobbling bands for $^{135}$Pr suggests that the PRM calculation
describes well the wave functions underlying the experimental
states. Let us now investigate the angular momentum geometry of the system
in detail.

\begin{figure*}[!ht]
  \begin{center}
    \includegraphics[width=14.0 cm]{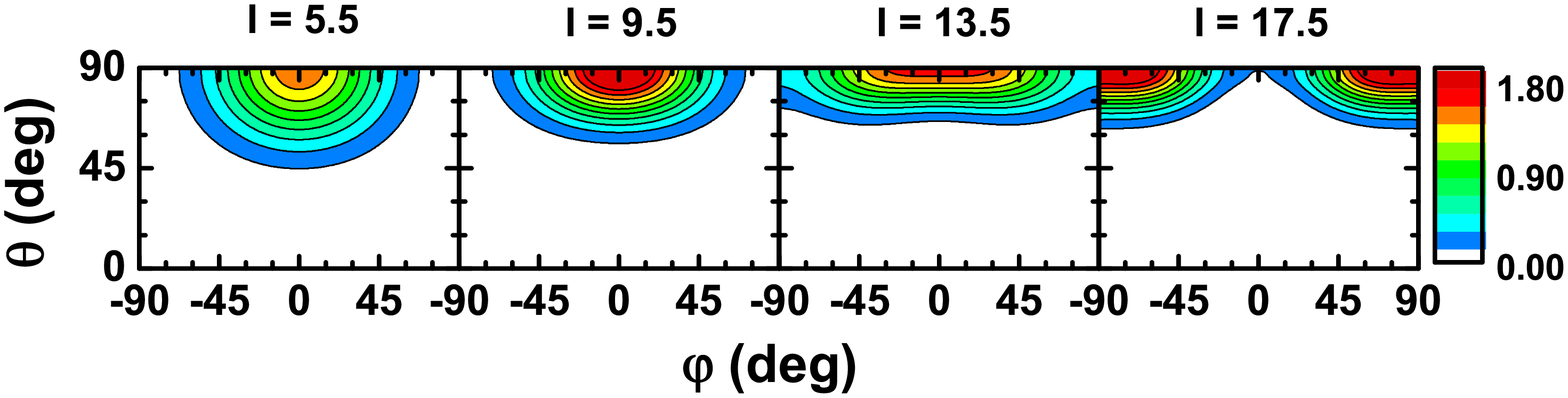}\\
    \includegraphics[width=14.0 cm]{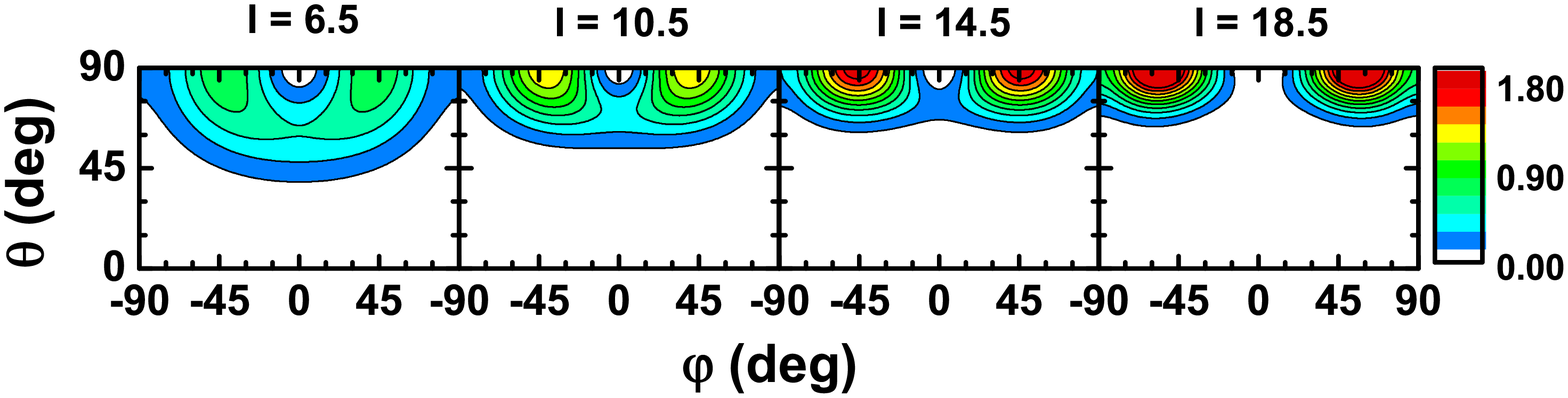}
    \caption{Azimuthal plots (i.e. distributions of the orientation of the angular momentum)
    calculated at $I=5.5$, 9.5, 13.5, and $17.5\hbar$ for the yrast band and at $I=6.5$, 10.5,
    14.5, and $18.5\hbar$ for the wobbling band in $^{135}$Pr.}\label{fig30}
  \end{center}
\end{figure*}

In Fig.~\ref{fig30}, the obtained profiles $\mathcal{P}(\theta,\varphi)$ for the
orientation of the angular momenta $\bm{I}$ in the $\theta\varphi$-plane
are shown at spin $I=5.5$, 9.5, 13.5, and $17.5\hbar$ for the yrast band, and
at $I=6.5$, 10.5, 14.5, and $18.5\hbar$ for the wobbling band in $^{135}$Pr.
We remind that $\theta$ is the angle between the $\bm{I}$ and the $l$-axis,
and $\varphi$ is the angle between the projection of $\bm{I}$ onto the $si$-plane
and the $s$-axis.

One observes that the maximum of $\mathcal{P}(\theta,\varphi)$ is always
located at $\theta=90^\circ$. This is because the $l$-axis carries the smallest MoI,
and in order to lower the energy the angular momentum prefers to lie in the
$si$-plane. Note that due to the $\textrm{D}_2$ symmetry,
$\mathcal{P}(\theta,\varphi)$ is an even function of $\varphi$.
For the states in the yrast band, the $\varphi$-coordinates
of the maxima gradually deviate from zero with increasing spin.
As a result, the number of maxima changes from one to two.
This implies that the rotational mode in the yrast band changes from a
principal axis rotation at the low spins ($I=5.5$ and $9.5\hbar$) to a
planar rotation at high spins ($I=13.5\hbar$). By examining
the profiles $\mathcal{P}(\theta,\varphi)$ for all yrast states, we
find that $I=13.5\hbar$ is the critical spin at which the rotational mode
changes (with $\varphi \simeq \pm 5^\circ$ at the maxima). At $I=17.5\hbar$,
the $\varphi$-coordinates of the maxima of $\mathcal{P}(\theta,\varphi)$
approaches $\pm 90^\circ$. In this case, the rotational mode changes
from a planar rotation back to a principal axis rotation about the $i$-axis.
These features are similar to the behavior of the minima of the total Routhian
surface as a function of the rotational frequency, calculated
by TAC in the Refs.~\cite{Q.B.Chen2014PRC, Q.B.Chen2016PRC_v1}. Both PRM and TAC
present the same physics picture: a principal axis rotation about the $s$-axis at
low spins, a transition to planar rotation at intermediate spins, and
a return to principal axis rotation about the $i$-axis at high spins.

In the lower part of Fig.~\ref{fig30}, the distributions $\mathcal{P}(\theta,\varphi)$
exhibit a different behavior in the wobbling band. With one-phonon excitation (wobbling
motion), the profiles $\mathcal{P}(\theta,\varphi)$ have two maxima for all spins.
At low spins ($I\leq 12.5\hbar$), the excitation is transverse wobbling about the
$s$-axis. This is reflected by the larger $\varphi$-values of the maxima of
$\mathcal{P}(\theta,\varphi)$ in wobbling states (with spin $I$) compared
to those of the corresponding yrast states (with spin $I-1$). Note that for the
zero-phonon states (with $I\leq 11.5\hbar$) the underlying wave functions are symmetric
and peaked at $\varphi=0^\circ$ ($s$-axis), whereas for one-phonon states
($I=6.5\hbar$, $8.5\hbar$, etc.) they are antisymmetric and have a node at $\varphi=0^\circ$. At
high spins ($I\geq 17.5\hbar$), the excitation from the yrast band into the wobbling
band is longitudinal wobbling about the $i$-axis. This is in accordance with the fact
that the $\varphi$-coordinate of the maxima of $\mathcal{P}(\theta,\varphi)$ in
the wobbling states (with spin $I$) are smaller than those in
the yrast states (with spin $I-1$). Moreover, the zero-phonon
state ($I=17.5\hbar$) is peaked at $\varphi=\pm 90^\circ$ ($i$-axis),
while the one-phonon state ($I=18.5\hbar$) has a node there. These features
are similar to the properties obtained with wave functions calculated from
a collective Hamiltonian in Refs.~\cite{Q.B.Chen2014PRC, Q.B.Chen2016PRC_v1}.

Therefore, we have confirmed that with the increasing spin, the wobbling mode
varies from the transverse at low spins to longitudinal at high spins,
which is consistent with the evolution of the wobbling frequency in Fig.~\ref{fig10}.
In fact, this variation is mainly driven by the collective rotor (cf. Fig.~\ref{fig80}).

\subsection{$R$-plots}

According to the above analysis, the collective rotor plays an essential role
in the wobbling motion. Therefore, we investigate in the following the probability
distribution of the rotor angular momentum ($R$-plots) as well as the its projections
onto each principal axis ($K_R$-plots).

\begin{figure}[!ht]
  \begin{center}
    \includegraphics[width=12.5 cm]{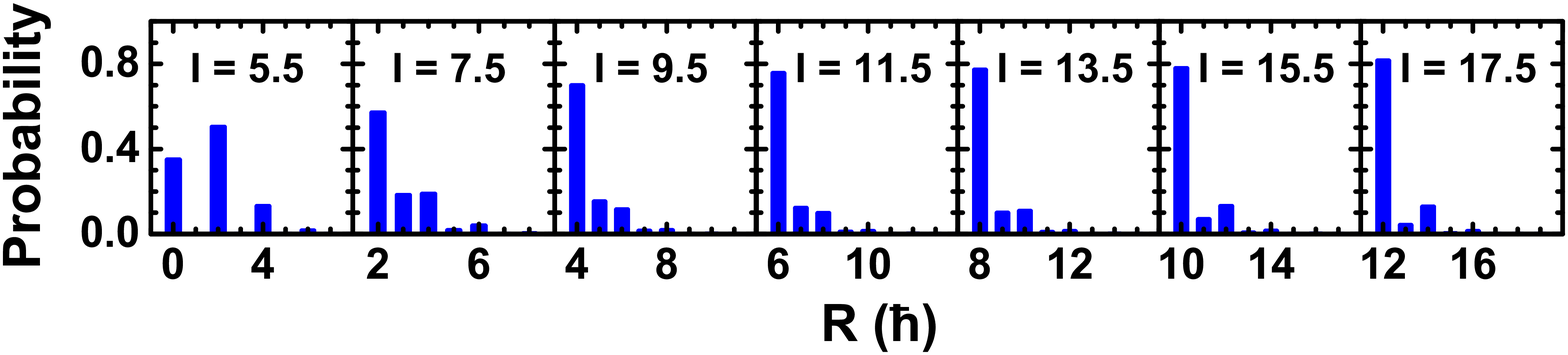}\\
    \includegraphics[width=12.5 cm]{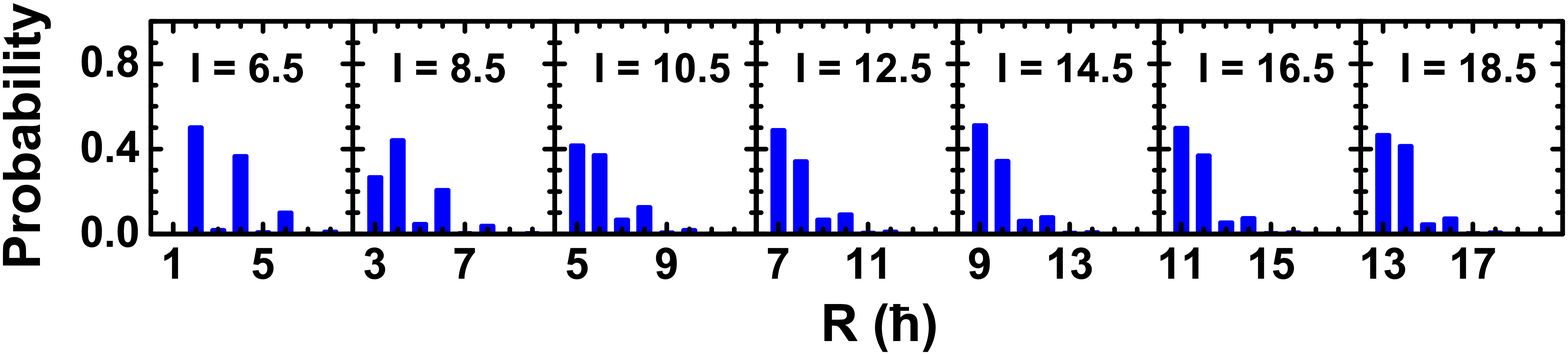}
    \caption{The probability distributions for the angular momentum of rotor
    ($R$-plots) for the yrast and wobbling bands in $^{135}$Pr.}\label{fig40}
  \end{center}
\end{figure}

In Fig.~\ref{fig40}, the probability distributions $P_{R}$ of the rotor angular
momentum ($R$-plots) calculated by Eq.~(\ref{eq9}) are displayed for the yrast and
wobbling bands in $^{135}$Pr. For a given spin $I$, the integer $R$ takes
values from $|I-j|$ to $I+j$, excluding $R=1$. It is found that for all $I$
the probability $P_R$ almost vanishes for large $R$. Therefore, the $R$-plots
are restricted in Fig.~\ref{fig40} to small $R$.

For the yrast band, $P_R$ has a pronounced peak at $R_{\textrm{min}}=|I-j|$,
except for $I=5.5\hbar$ (the bandhead), where the maximal weight occurs at
$R=|I-j|+2=2\hbar$. For the wobbling band, $P_R$ has two peaks of similar
height, which are located at $R=|I-j|$ and $|I-j|+1$. An exception is again
the bandhead $I=6.5\hbar$, where the peaks lie at $R=|I-j|+1$ and $|I-j|+3$.
The $R$-plots indicate that $R$ is an asymptotic good quantum number in
the yrast band ($I\geq 7.5\hbar$), but not in the wobbling band. This is
different to the wobbling motion of a pure triaxial rotor, where $R$ is
a good quantum in all bands~\cite{Bohr1975, W.X.Shi2015CPC}. However,
it should be noted that the admixture of the states with $R=|I-j|$
and $R=|I-j|+1$ in the wobbling band is important as it provides
the possibility for the (quantum mechanical) wobbling transition.
This admixture causes that the average value of $R$ in the wobbling band
$R_{\textrm{wobb}}(I)$ at spin $I$ is larger than $|I-j|$ and leads to
$R_{\textrm{wobb}}(I)-R_{\textrm{yrast}}(I-1)>1\hbar$,
so that the rotor in the wobbling band with spin $I$ has to wobble
to increase its spin by only $1\hbar$ with respect to the yrast band
(with spin $I-1$).

\subsection{$K_R$-plots}

In the following the probability distributions for the projections ($K_R=R_l$, $R_s$,
and $R_i$) of the rotor angular momentum onto the $l$-, $s$-, and $i$-axes ($K_R$-plots)
will be investigated. For the triaxiality parameter
$\gamma=-26^\circ$, the $l$-axis is the designated quantization axis. The distributions
with respect to the $s$- and $i$-axis are obtained by taking $\gamma=146^\circ$
and $266^\circ$, respectively. These $\gamma$-values correspond to the equivalent sectors
such that the nuclear shape remains the same, but only the principal axes are
interchanged~\cite{Ring1980book}.

\begin{figure}[!ht]
  \begin{center}
    \includegraphics[width=12.5 cm]{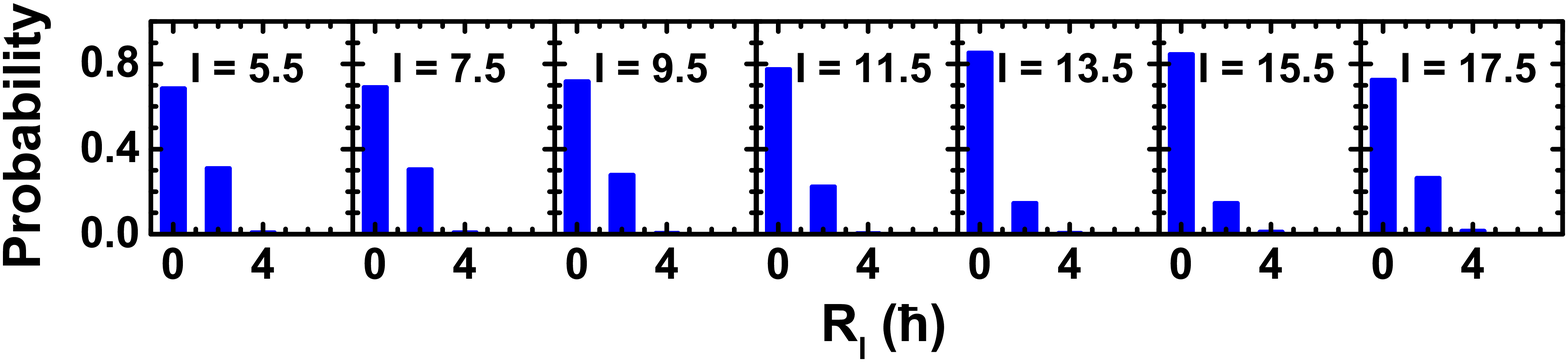}\\
    \includegraphics[width=12.5 cm]{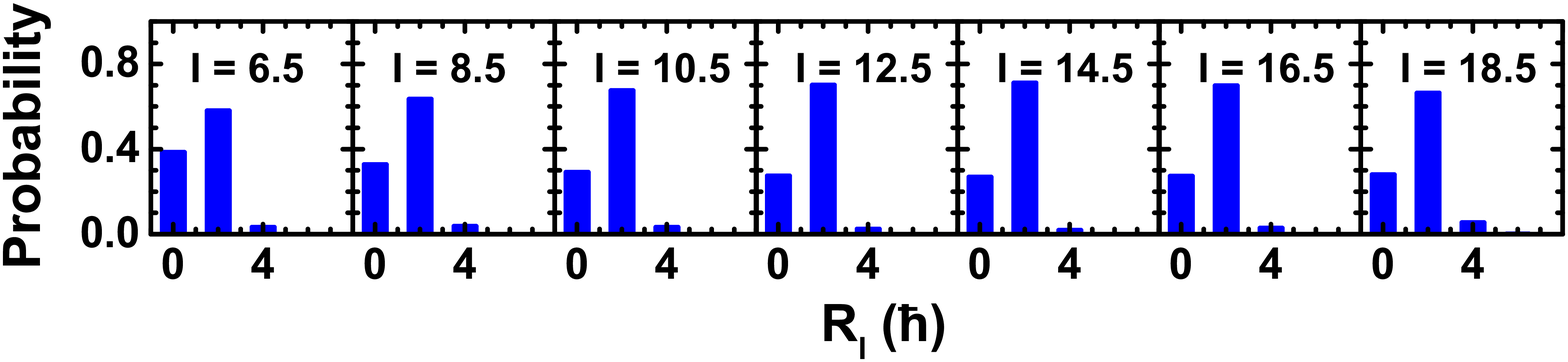}
    \caption{Probability distributions for the projection of the rotor angular momentum
    onto the $l$-axis for the yrast and wobbling bands in $^{135}$Pr.}\label{fig70}
  \end{center}
\end{figure}

In Fig.~\ref{fig70}, the probability distributions for the projection
of the rotor angular momentum onto the $l$-axis $P_{R_l}$ as calculated
in the PRM, are shown for the yrast and wobbling bands in $^{135}$Pr.
For both the yrast and wobbling bands, $P_{R_l}$ has two peaks at $R_l=0$
and $2\hbar$, indicating that the rotor angular momentum has only very
small components along the $l$-axis, to which a very small MoI is associated.
This is consistent with the azimuthal plots shown in Fig.~\ref{fig30}. At the same
time, the distributions of $P_{R_l}$ for the yrast and the wobbling bands do
not change much as the spin $I$ increases, indicating that the rotor angular
momentum component along the $l$-axis remains almost constant. For the yrast
band, $P_{R_l}$ at $R_l=0\hbar$ is much larger than $P_{R_l}$ at $R_l=2\hbar$,
while for the wobbling band, the situation is opposite. There, $P_{R_l}$ at
$R_l=2\hbar$ is larger than $P_{R_l}$ at $R_l=0\hbar$.

\begin{figure}[!ht]
  \begin{center}
    \includegraphics[width=12.5 cm]{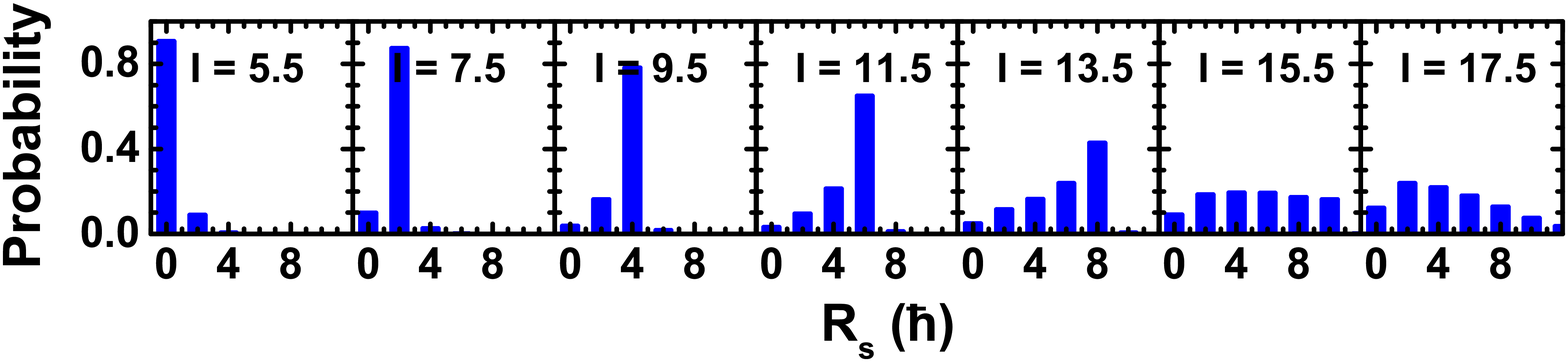}\\
    \includegraphics[width=12.5 cm]{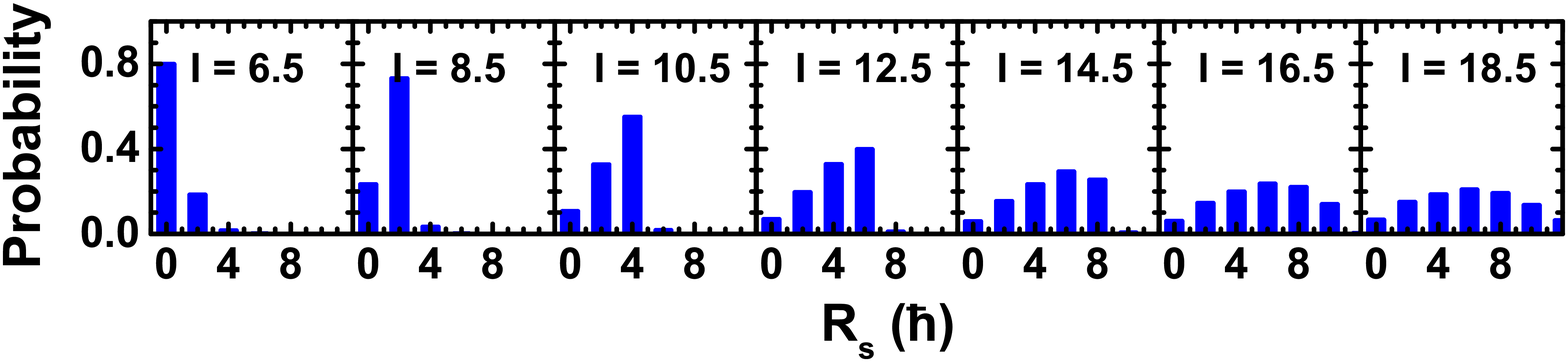}
    \caption{Same as Fig.~\ref{fig70}, but for the projection onto the $s$-axis.}\label{fig60}
  \end{center}
\end{figure}

The probability distributions $P_{R_s}$ of the component $R_s$ are displayed
in Fig.~\ref{fig60} for the yrast and wobbling bands in $^{135}$Pr.
In the region $I\leq 13.5\hbar$, the distributions $P_{R_s}$ for states
in the yrast band (with $I-1$) and the wobbling band (with $I$) show a similar
behavior. This indicates that the rotor angular momenta of states in the
yrast (with $I-1$) and wobbling (with $I$) bands have similar components
along the $s$-axis due to the transverse wobbling motion. For
neighboring states with $I-2$ and $I$, the distance between the
peaks of $P_{R_s}$ is $2\hbar$. In the region $I\geq 14.5\hbar$,
where the transverse wobbling motion disappears, the distributions
$P_{R_s}$ are spread over many $R_s$-values. The average value of $R_s$
is about $4\hbar$ for the yrast band and about $6\hbar$ for the wobbling
band.

\begin{figure}[!ht]
  \begin{center}
    \includegraphics[width=12.5 cm]{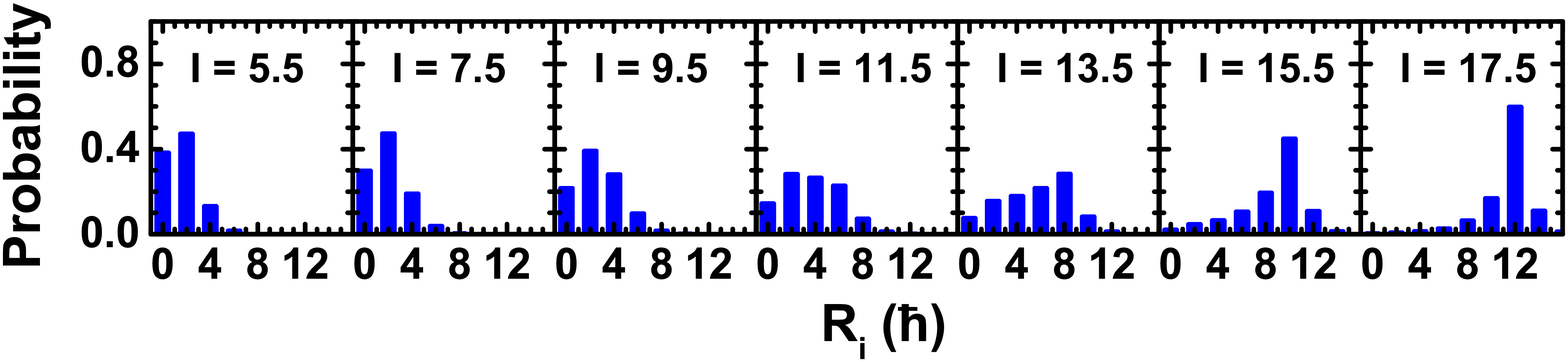}\\
    \includegraphics[width=12.5 cm]{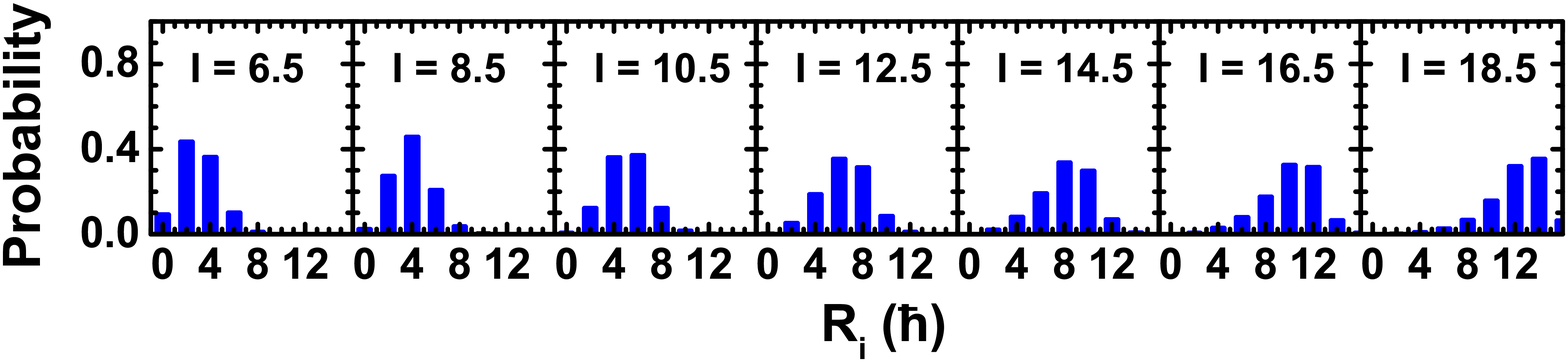}
    \caption{Same as Fig.~\ref{fig70}, but for the projection onto the $i$-axis.}\label{fig50}
  \end{center}
\end{figure}

In Fig.~\ref{fig50}, the probability distributions $P_{R_i}$ of the component
$R_i$ are shown for the yrast and wobbling bands in $^{135}$Pr. In comparison
to $P_{R_l}$ and $P_{R_s}$, the distributions $P_{R_i}$ reveal stronger
admixtures of the various values of $R_i$, which originates from the wobbling
motion of the rotor towards the $i$-axis. One also observes that $P_{R_i}$ of
the yrast and wobbling bands behavior differently. In the region $I\leq 13.5\hbar$,
the probability $P_{R_i}$ at $R_i=0\hbar$ has a finite value in the yrast band,
while it vanishes for the wobbling band. This is a characteristic of the one-phonon
excitation of the wobbling motion. Namely, the underlying wave function for a
zero-phonon state (yrast band) is even under $R_i \to -R_i$, whereas for
a one-phonon state (wobbling band) it is odd. This picture is
also consistent with the features displayed in the azimuthal plots
(cf. Fig.~\ref{fig30}). The peak position of the distribution $P_{R_i}$
increases by about $2\hbar$ from a state in the yrast band (with $I-1$) to a state
in the wobbling band (with $I$). This increment is caused by the wobbling
motion from the $s$-axis towards the $i$-axis. For neighboring states
with $I-2$ and $I$, the average value of $R_i$ differs by about $1\hbar$.
This means that $R_i$ for the state $I$ in the yrast band is about $1\hbar$
smaller than for the state $I-1$ in the wobbling band.

In the region $I\geq 14.5\hbar$, the distributions $P_{R_i}$ for the yrast band
show a clear peak at $R_i=|I-j|=R$ (cf. Figs.~\ref{fig40} and \ref{fig50}),
indicating that the rotor has aligned with the $i$-axis. For the wobbling band
one observes two peaks of similar height at $R_i=|I-j|-1$ and
$|I-j|+1$, which gives an increment of $R_i$ by about $1\hbar$ from
the yrast state (with $I-1$) to the wobbling state (with $I$). This behavior
is different from the transverse wobbling region, where the increment is
about $2\hbar$.

\subsection{Angular momentum coupling schemes}

\begin{figure}[!ht]
  \begin{center}
    \includegraphics[width=6.0 cm]{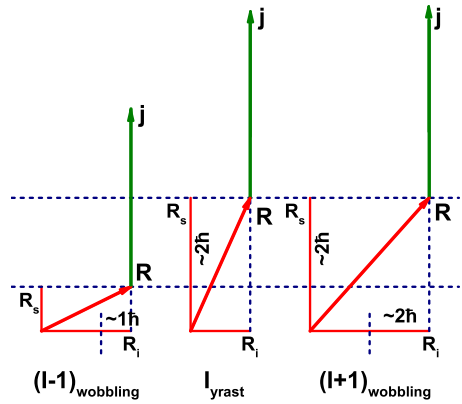}
    \caption{Schematic illustration of the coupling scheme of the angular
    momenta $\bm{j}$ and $\bm{R}$ of the high-$j$ particle and the rotor for the
    transverse wobbling in an yrast state with $I$ and two wobbling states
    with $I\pm 1$. The total angular momentum is
    $\bm{I}=\bm{R}+\bm{j}$.}\label{fig90}
  \end{center}
\end{figure}

From the above analysis of energy expectation values of the intrinsic
Hamiltonian $\hat{H}_{\textrm{intr}}$, azimuthal plots $\mathcal{P}(\theta,\varphi)$
of the total angular momentum, and the $R$-plots and three $K_R$-plots for
the rotor angular momentum, one can deduce the following features in the transverse
wobbling region:
\begin{itemize}
  \item[(i)] the single-particle (angular momentum) is aligned with the $s$-axis;
  \item[(ii)] the average rotor angular momentum is more than $1\hbar$ (and less than $2\hbar$)
  longer in the wobbling band with spin $I+1$ than in the yrast band with spin $I$;
  \item[(iii)] the projection of the rotor angular momentum onto the $l$-axis
  is very small;
  \item[(iv)] the rotor angular momenta in yrast states (with $I$)
  and wobbling states (with $I+1$) have similar components along the $s$-axis.
  For neighboring states with $I-2$ and $I$, the component $R_s$ differs by
  about $2\hbar$;
  \item [(v)] the component $R_i$ increases by about $2\hbar$ from an yrast state
  $I$ to a wobbling state $I+1$. In addition, $R_i$ in the yrast state $I$
  is about $1\hbar$ smaller than its value in the wobbling state $I-1$.
\end{itemize}
Combining these features, a schematic illustration of the coupling scheme of
the angular momenta $\bm{j}$ and $\bm{R}$, of the high-$j$ particle and the rotor,
for transverse wobbling in an yrast state $I$ and two wobbling states $I\pm 1$
is shown in Fig.~\ref{fig90}.

\begin{figure}[!ht]
  \begin{center}
    \includegraphics[width=10.0 cm]{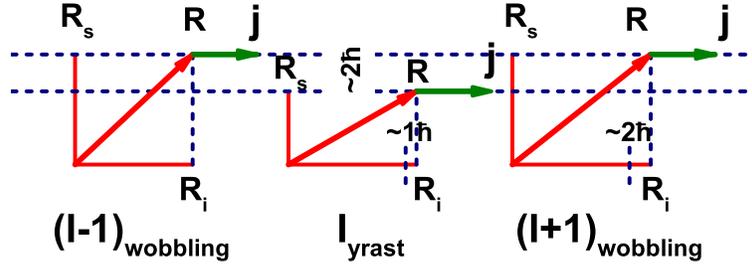}
    \caption{Similar as Fig.~\ref{fig90}, but for the longitudinal wobbling
    motion.}\label{fig20}
  \end{center}
\end{figure}

On the other hand, for longitudinal wobbling one finds the following features:
\begin{itemize}
  \item [(i)] the proton particle (angular momentum) is aligned with the $i$-axis;
  \item [(ii)] the average value of $R_s$ is about $4\hbar$ in the yrast band
  and about $6\hbar$ in the wobbling band.
  \item [(iii)] the increment of $R_i$ from an yrast state with
  $I-1$ to a wobbling state with $I$ is about $1\hbar$.
\end{itemize}
Again combining these features, a schematic illustration of the
coupling scheme of $\bm{j}$ and $\bm{R}$ for the longitudinal
wobbling motion in an yrast state with $I$ and two wobbling states
with $I\pm 1$ is shown in Fig.~\ref{fig20}. This coupling scheme
differs from that for transverse wobbling, shown in
Fig.~\ref{fig90}. One can clearly see that the rotor angular
momentum is much longer than the single particle angular momentum.
It should be noted that a schematic illustration of the longitudinal
wobbling motion has also been given in Refs.~\cite{Odegaard2001PRL,
Hamamoto2002PRC}, but there the MoI belonging to $s$-axis was
assumed to be the largest. In that case, the angular momenta of the
rotor and the particle both align with the $s$-axis in the yrast
band.


\section{Summary}\label{sec4}

In summary, the behavior of the collective rotor for the wobbling motion of
$^{135}$Pr has been investigated in the PRM. After successful reproduction
of the experimental energy spectra and the wobbling frequencies, the
separate contributions from the rotor and the single-particle Hamiltonian
to the wobbling frequencies have been analyzed. It is found that the
collective rotor motion is responsible for the decrease of the wobbling
frequency in transverse wobbling, and its increase
in longitudinal wobbling.

The evolution of the wobbling mode in $^{135}$Pr from transverse
at low spins to longitudinal at high spins has been illustrated by
the distributions $\mathcal{P}(\theta,\varphi)$ of the total angular momentum
in the intrinsic frame (azimuthal plots). According to the analysis of the
probability distributions of the rotor angular momentum ($R$-plots)
and their projections onto the three principal axes
($K_R$-plots), different schematic coupling schemes of the
angular momenta $\bm{j}$ and $\bm{R}$ of the rotor and the high-$j$ particle
in the transverse and longitudinal wobbling have been obtained.

In perspective, the $R$-plots and $K_R$-plots presented in this work can
be used to examine the fingerprints of electromagnetic transitions ($E2$
or $M1$) between wobbling bands, or can be extended to investigate, e.g.,
the behavior of the collective rotor for chiral rotation~\cite{Frauendorf1997NPA}.

\section*{Acknowledgements}

One of the authors (Q.B.C.) thanks S. Frauendorf for helpful discussions.
Financial support for this work was provided by Deutsche
Forschungsgemeinschaft (DFG) and National Natural Science Foundation
of China (NSFC) through funds provided to the Sino-German CRC 110
``Symmetries and the Emergence of Structure in QCD''. The work of UGM
was also supported by the Chinese Academy of Sciences (CAS) President's
International Fellowship Initiative (PIFI) (Grant No. 2018DM0034) and
by VolkswagenStiftung (Grant No. 93562).


\end{CJK}

\end{document}